\documentstyle [12pt]{article}
\newcommand{\bce}{\begin{center}}
\newcommand{\ece}{\end{center}}
\newcommand{\beq}{\begin{equation}}
\newcommand{\eeq}{\end{equation}}
\newcommand{\bef}{\begin{figure}}
\newcommand{\eef}{\end{figure}}
\newcommand{\bea}{\vspace{0.25cm}\begin{eqnarray}}
\newcommand{\eea}{\end{eqnarray}}

\newcommand{\ba}{\begin{array}}
\newcommand{\ea}{\end{array}}

\newcommand{\singlespace}{
    \renewcommand{\baselinestretch}{1}\large\normalsize}
\newcommand{\doublespace}{
    \renewcommand{\baselinestretch}{1.6}\large\normalsize}

\newcommand{\footju}{\footnote[1]{ also: IKP (Theorie),
Forschungszentrum J\"ulich, D-52425 J\"ulich, F.R.G.}}
\newcommand{\footbo}{\footnote[2]{ also: ITKP, Universit\"at Bonn,
D-53115 Bonn, F.R.G.}}
\begin{document}
\pagestyle{empty}
\singlespace
\vspace{1.0in}
\begin{flushright}
January, 1995 \\
\end{flushright}
\vspace{1.0in}
\bce
{\large{\bf Thermal Properties of a Hot Pion Gas beyond
the Quasiparticle Approximation}}
\vskip 1.0cm
R. Rapp\footju and J. Wambach{\footnotemark[1] \footbo}

\vspace{.35in}
{\it Department of Physics\\
University of Illinois at Urbana-Champaign\\
1110 West Green St., Urbana, IL 61801-3080, USA}
\ece
\vspace{.65in}
\begin{abstract}
\noindent
Within the Matsubara formalism we derive expressions for the pion
selfenergy and the two-pion propagator in a hot pion gas. These
quantities are used to selfconsistently calculate the
in-medium $\pi\pi$ amplitude beyond the
quasiparticle approximation (QPA). The results are shown to differ
significantly from QPA-based calculations. We also examine the impact
of chiral constraints on the $\pi\pi$ interaction in a chirally
improved version of the J\"ulich $\pi\pi$ model.
\end{abstract}
\vspace{.75in}
\begin{flushright}
\singlespace
PACS Indices: 13.75.Lb\\
13.85.-t\\  14.40.aq
\end{flushright}
\newpage
\pagestyle{plain}
\baselineskip 16pt
\vskip 48pt

\newpage
\doublespace

\section{Introduction}
\noindent It is anticipated that future experiments at the Brookhaven
Relativistic Heavy Ion Collider (RHIC) will create a highly
excited, but nearly baryon-free central zone.
At later stages of the collision
this zone will contain a dense meson gas, primarily consisting
of pions with temperatures of the order of the pion mass.
The equation of state and the transport properties of the
pion gas are therefore of great current interest~\cite{BuKa,Gavi}.
One expects that the $\pi\pi$ interaction will be modified due to
statistical correlations (Bose occupation factors)
as well as dynamical (selfenergy)
effects~\cite{Shur,ACSW,Sche,BBDS,ChDa}.
In this context we have previously  presented a
Brueckner-type calculation for a thermal pion gas
treating both effects selfconsistently~\cite{RaWa}.
To facilitate  the numerical calculations we have employed
the quasiparticle approximation (QPA) for the single-pion propagator.
We have noticed, however, that this approximation
becomes unreliable for temperatures above $T\approx 150$ MeV,
{\it i.e.} in the vicinity of $T_c$, the chiral restoration
temperature. In addition, recent studies of pion correlations
in cold nuclear matter~\cite{ARCSW} have shown that, for
a thorough description of the subthreshold region, it is
mandatory to work with a $\pi\pi$ model which obeys
constraints imposed by chiral symmetry. Since chiral symmetry
predicts properties of the $\pi\pi$ interaction in the
soft pion limit ($m_\pi \rightarrow$0), there is an
intimate relation between the proper off-shell treatment of in-medium
properties and the implementation of chiral constraints.

\noindent The aim of the present paper is twofold: first we
investigate effects of proper off-shell calculations
by comparing them to the QPA results of
ref.~\cite{RaWa}. This will be discussed in sect.~2.
To incorporate constraints from chiral symmetry we then present in
sect.~3 an improved version of the $\pi\pi$
interaction of Lohse et al.~\cite{LDHS,PeHS}
which is subsequently used for the in-medium
calculations. A short summary and concluding
remarks are given in sect.~4.

\section{Formalism and Results from the  J\"ulich $\pi\pi$ Model}
In the $\pi\pi$ meson exchange model of Lohse et al.~\cite{LDHS,PeHS}
the vacuum $\pi\pi$ T-matrix in a given partial-wave/isospin channel,
$JI$, is obtained from a Lippmann-Schwinger equation, where the
Blankenbecler Sugar reduction scheme~\cite{BbSu} of
the 4-dimensional Bethe Salpeter equation is adopted:
\bea
T_{\pi\pi}^{JI}(E,q_1,q_2) & = & V_{\pi\pi}^{JI}(E,q_1,q_2)
     \nonumber\\
 & &
+ \int\limits_0^{\infty}dk \ k^2 \ 4{\omega}_k^2 \
V_{\pi\pi}^{JI}(E,q_1,k)
 \   G_{\pi\pi}(E,k) \ T_{\pi\pi}^{JI}(E,k,q_2)  \ .
\label{eq:Tmat}
\eea
Here $E$ denotes the starting energy of the
pion pair and $q_1$, $q_2$ are the
relative CMS 3-momenta of one pion in the in- and outgoing state,
respectively. The interaction kernel $V^{JI}_{\pi\pi}$ is calculated
from an effective meson Lagrangian including $\rho$-,
$f_0$(1400)- and $f_2$(1270)-s-channel exchanges, a $\rho$-meson
exchange in the t-channel as well as
coupling to the $K\bar K$ channel (not written explicitely
in eq.~(1)). The vacuum BbS two-pion propagator in the CMS reads
\bea
G_{\pi\pi}^0(E,k) & = & \int \frac{i \ dk_0}{2\pi} \ D_\pi(k_0,k) \
D_\pi(E-k_0,k)  \nonumber \\
 & = & \frac{1}{\omega_k} \  \frac{1}{E^2-4\omega_k^2+i\eta}  \ ,
\eea
where $D_\pi(k_0,k)$ denotes the single-pion  propagator and
$\omega_k=m_\pi^2+k^2$.

\noindent
At moderate gas densities it is sufficient to treat medium
modifications at the one- and two-body level rendering a
Brueckner-type description. Leaving the pseudopotentials
$V_{\pi\pi}^{JI}$ temperature independent, which should be
a good approximation for temperatures well below the masses of
the exchanged mesons, the temperature dependence of the T-matrix
is solely induced by the thermal two-pion propagator,
$G_{\pi\pi}^T$. In terms of the thermal single-pion propagator,
$D_\pi$, it can be
evaluated by means of a Matsubara sum~\cite{KaBa,FeWa}:
\beq
G_{\pi\pi}^T(\Omega_\lambda,\vec k_1, \vec k_2)= T \
\sum_{z_\nu} D_\pi(z_\nu,\vec k_1) \ D_\pi(\Omega_\lambda
-z_\nu, \vec k_2) \ ,
\eeq
with discrete frequencies
\beq
z_\nu=i\omega_\nu, \ \Omega_\lambda=i\omega_\lambda, \
\omega_\nu=2\pi\nu T \  , \ \nu, \lambda \ {\rm integer}.
\eeq
Using standard techniques of analytical continuation
one then obtains
\bea
G_{\pi\pi}^T(E_+,\vec k_1,\vec k_2) = \frac{1}{2} \int
\frac{d\omega}{\pi}  \frac{d\omega '}{\pi}
Im D_\pi(\omega_+,\vec k_1) \
Im D_\pi(\omega '_+,\vec k_2)
\ (1+f^\pi(\omega)+f^\pi(\omega '))
\nonumber \\  \hspace{1cm}
\times  \lbrack \frac{1}{\omega+\omega '-E_-}
+\frac{1}{\omega+\omega '+E_+}
\rbrack \
\eea
where $\omega_{\pm}=\omega\pm i\eta $ {\it etc.} and
$f^\pi(\omega)= (\exp \lbrack \omega /T \rbrack -1)^{-1}$.
For the T-matrix evaluation it will be sufficient to consider
$G_{\pi\pi}^T$ for E$>$0 and in the two-pion CMS, {\it i.e.}
in back-to-back kinematics with
$\vec k_2$=--$\vec k_1$. In this case the imaginary part
of eq.~(5) receives 2 contributions:
\bea
Im G_{\pi\pi}^{T,A}(E_+,k)= -\int\limits_0^E \frac{d\omega}{\pi}
\lbrack 1+f^\pi(\omega)+f^\pi(E-\omega)\rbrack
Im D_\pi(\omega_+,k) Im D_\pi(E-\omega_-,k)  ,  \nonumber \\
Im G_{\pi\pi}^{T,B}(E_+,k)=-\int\limits_{E}^\infty \frac{2d\omega}{\pi}
\lbrack f^\pi(\omega-E)-f^\pi(\omega)\rbrack Im D_\pi(\omega_+,k)
Im D_\pi(\omega_+-E,k)  \ \ .  \ \
\eea
In the zero-temperature limit contribution A reproduces
the imaginary part of eq.~(2), whereas B vanishes.
The latter arises solely from thermal excitations.
The real part of $G_{\pi\pi}^T$ is calculated via a dispersion
integral~\cite{ARCSW}:
\beq
Re G_{\pi\pi}^T(E,k)=-{\cal P}\int\limits_0^{\infty}
\frac{d{E'}^2}{\pi} \frac{Im G_{\pi\pi}^T(E'_+,k)}{E^2-{E'}^2}  \  .
\eeq

\noindent
In a second step we evaluate the pion selfenergy, $\Sigma_\pi$,
which enters the single-pion propagator
\beq
D_\pi(\omega_+,k)=\frac{1}{(\omega_+)^2-m_\pi^2-k^2
-\Sigma_\pi(\omega_+,k)} \ .
\eeq
At the two-body level $\Sigma_\pi$ is expressed in terms of
the $\pi\pi$ invariant forward scattering amplitude as
\beq
\Sigma_\pi(z_\nu,\vec k)=-\int\frac{d^3p}{(2\pi)^3}  \ T \
\sum_{z_{\nu '}} M_{\pi\pi}(z_\nu+z_{\nu '},\vec k,\vec p)
D_\pi(z_{\nu '},\vec p) \  ,
\eeq
which involves a Matsubara sum over $z_{\nu '}$.
Using the spectral representations of $M_{\pi\pi}$  and $D_\pi$
one arrives at~\cite{SRS}
\beq
\Sigma_\pi(\omega_+,\vec k)=\int\frac{d^3p}{(2\pi)^3}
\frac{dE'}{\pi}\frac{d\omega '}{\pi} Im M_{\pi\pi}(E'_+,
\vec k,\vec p) Im D_\pi(\omega'_+,\vec p)
\frac{f^\pi(\omega ')-f^{\pi}(E')}{\omega_++\omega '-E'} .
\eeq
The imaginary part of $\Sigma_\pi$ is given by the
$\delta$- function contribution of $1/(\omega_++\omega'-E')$.
The positive-energy part of the $\omega '$ integration gives
rise to the standard contribution commonly used in the
literature:
\beq
Im \Sigma_\pi^A(\omega_+,k)=-\int\frac{d^3p}{(2\pi)^3}
\int\limits_{0}^{\infty}
\frac{d\omega '}{\pi} Im M_{\pi\pi}(\omega_++\omega ',
\vec k, \vec p) Im D_\pi(\omega_+',\vec p)
\lbrack f^\pi(\omega ')-f^\pi(\omega+\omega ')\rbrack \ .
\eeq
Note that the Matsubara formalism yields the 'correction'
$f^\pi(\omega+\omega ')$ in the occupation factor as was found by
Chanfray et al.~\cite{ChDa} in the real-time formalism.
The negative-energy part of the $\omega '$ integration can be
rewritten by exploiting the symmetry properties of the 1-pion
propagator, the $M_{\pi\pi}$-amplitude and the Bosefactor
resulting in
\bea
Im \Sigma_\pi^B(\omega_+,k)=-\int\frac{d^3p}{(2\pi)^3}
\int\limits_{0}^{\infty}
\frac{d\omega '}{\pi} Im M_{\pi\pi}(|\omega -\omega '|_+,
\vec k, \vec p) Im D_\pi(\omega_+',\vec p) \nonumber\\
\lbrack {\rm sgn}(\omega -\omega ') f^\pi(\omega ')-
f^\pi(|\omega-\omega '|)\rbrack \ .
\eea
with sgn$(\omega -\omega ')=\pm 1$. In analogy to the thermal
two-pion propagator (eq.~(6)) the contribution B accounts
for thermal excitations of the system. E.g. in lowest order,
$Im\Sigma_\pi^B$ is only non-vanishing for $\omega > 3m_\pi$.
The $M_{\pi\pi}$-amplitude is related to the $T_{\pi\pi}$-matrix after
transforming it into the CMS and identifying the starting energy
as $E^2\equiv s=(\omega\pm\omega ')^2-(\vec p +\vec k)^2$:  \\
\beq
M_{\pi\pi}(\omega\pm\omega ',\vec k,\vec p)=M_{\pi\pi}(\sqrt s,q,q)
=(2\pi)^3 \ 4\omega_q^2 \ T_{\pi\pi}(E,q,q)
\eeq
with $q=q(\omega,\omega ',k,p,\cos \Theta)$ being the CMS momentum
of both in- and outgoing pions. After a variable transformation
$\cos\Theta\rightarrow E$, where $\Theta =\angle (\vec k,\vec p)$,
the final expressions read
\bea
Im\Sigma_\pi(\omega_+,k) & = & Im\Sigma_\pi^A(\omega_+,k)+
Im\Sigma_\pi^B(\omega_+,k) \nonumber\\
Im\Sigma_\pi^A(\omega_+,k) & = & -\frac{1}{k} \int\limits_0^\infty
\frac{p \ dp}{(2\pi)^2} \int\limits_0^\infty \frac{d\omega '}{\pi}
Im D_\pi(\omega_+',p) \lbrack f^\pi(\omega ')
-f^\pi(\omega+\omega ')\rbrack \ I^A \nonumber\\
Im\Sigma_\pi^B(\omega_+,k) & = & -\frac{1}{k} \int\limits_0^\infty
\frac{p \ dp}{(2\pi)^2} \int\limits_0^\infty \frac{d\omega '}{\pi}
Im D_\pi(\omega_+',p) \lbrack {\rm sgn}(\omega-\omega ')
f^\pi(\omega ') +f^\pi(|\omega-\omega '|)\rbrack \ I^B \nonumber\\
\eea
with
\beq
I^{A/B}=\int\limits_{E_{min}^{A/B}}^{E_{max}^{A/B}} dE \ E \
M_{\pi\pi}(E,q,q) \ ,
\eeq
where the upper and lower bounds,
\bea
E_{max/min}^A=(\omega +\omega ')^2-k^2-p^2\pm 2kp \ , \nonumber\\
E_{max/min}^B=(\omega -\omega ')^2-k^2-p^2\pm 2kp \ ,
\eea
correspond to $\cos\Theta=\mp 1$, respectively.
For the numerical evaluation of $Im M_{\pi\pi}$ we include
all partial waves up to $J=2$ and all isospins.
Eq.~(10) evades a 'direct' calculation of $Re \Sigma_\pi(\omega,k)$
in terms of $Re M_{\pi\pi}$ due to the $E'$-dependence of
the occupation factor. Therefore we again make use of a
dispersion relation including a subtraction at $\omega =0$:
\beq
Re \Sigma_\pi(\omega,k)=-{\cal P} \int_0^\infty \ \frac{d{\omega '}^2}
{\pi} \frac{Im \Sigma_\pi(\omega_+',k)}{\omega^2-{\omega '}^2}
\frac{\omega^2}{{\omega '}^2} \ .
\eeq
The subtraction is dictated by the fact that $Re \Sigma_\pi(\omega,
k)\rightarrow 0$ for $\omega\rightarrow 0$. To check the
subtraction procedure numerically we have performed calculations
by approximating $f^\pi(E')\approx f^\pi(\omega+\omega ')$ in
eq.~(10), which allows to reinsert the spectral representation
of $M_{\pi\pi}$.  $Re \Sigma_\pi$ can then be calculated directly
in terms of $Re M_{\pi\pi}$ (in exact analogy to calculating
$Im \Sigma_\pi$ in terms of $Im M_{\pi\pi}$). The
results of this test calculation coincide within a few percent
with the results obtained from eq.~(17).
Eqs.~(1),~(6)-(8) and~(14),~(17) form
the selfconsistent set of equations which is solved by
numerical iteration.

\noindent The results for the imaginary part of
the $\pi\pi$ M-amplitude
are displayed in Fig.~1. With increasing temperature
we find in the $\sigma$-channel ($JI$=00)
a considerable enhancement over the vacuum curve for CMS energies
below E$\approx$450~MeV.
Apart from the highest temperature (T=200~MeV) there is, however,
no significant strength below 2$m_\pi$.
For T=200~MeV some accumulation of subthreshold strength
is found which is almost entirely due to thermal excitations (the
'B'-contribution in eq.~(6)).
For comparison,  results within the
QPA are also given (in contrast to ref.~\cite{RaWa}
the 'polestrength'
$z_k=(1-\partial Re\Sigma_\pi(\omega,k)/\partial\omega^2|_{e_k})^{-1}$
is not included since it leads to a violation of the normalization
condition of the 1-pion spectral function).
For the $\rho$-channel ($JI$=11) both the full and the QPA results show
a broadening of the resonance, increasing with temperature due to the
thermal motion of the gas particles.
The QPA results clearly overestimate the thermal broadening and
the upward peakshift of the $\rho$ resonance. \\
A comment  concerning the QPA
calculations of refs.~\cite{RaWa,RW94} is in order.
When expanding around the quasiparticle pole, in next-to-leading
order a complex polestrength $z_k$ arises.
This results in a strong enhancement
of $Im M_{\pi\pi}$ in the vicinity of resonance peaks. Such a
strong enhancement is not
confirmed by the full calculations presented here.
At intermediate energies the full calculations do show
an enhancement over the QPA results, however.

\section{$\pi\pi$ Interaction with Chiral Constraints}
For the subthreshold $\pi\pi$ amplitude in cold
nuclear matter the
implementation of properties dictated by chiral symmetry
has turned out to be crucial~\cite{ARCSW}. To see the impact of
such constraints on the interacting pion gas we have improved
the J\"ulich model employed in the previous section in two respects
(for a detailed discussion see refs.~\cite{DRW,ARCSW}):
\begin{itemize}
\item[--] we introduce $\pi\pi$ contact interactions as obtained
from the gauged nonlinear $\sigma$ model of Weinberg~\cite{We66}.
At the tree level these contributions ensure the correct values
of the $\pi\pi$ amplitude at both the soft point and the Adler point.
When supplemented with phenomenological form factors, $F(q)=(2\Lambda^2
-4m_\pi^2)^2/(2\Lambda^2+4q^2)^2$, they are
suitable for iteration in the T-matrix equation~(1);
\item[--] instead of using the on-energy-shell prescription
implied by the BbS formalism  we here employ an on-mass-shell
prescription for the off-shell values
of the pseudopotentials $V^{JI}_{\pi\pi}$.
In this way we are able to impose the correct chiral limit of
the s-wave scattering lenghts, {\it i.e.} $a^{0I}\rightarrow 0$
for $m_\pi\rightarrow 0$ (the model of Lohse et al.~\cite{LDHS}
does not have this property).
\end{itemize}
With similar cutoff parameters as in \cite{LDHS} the results of
refitting the vacuum $\pi\pi$ scattering data are quite satisfactory
(see Fig.~2).
The additional cutoff parameters in the
contact terms are chosen to be slightly below 1 GeV.
For the s-wave scattering lengths we obtain $a^{00}=0.21m_\pi^{-1}$,
$a^{02}=-0.028m_\pi^{-1}$, in reasonable agreement with a
recent experimental analysis~\cite{BuLo}.

\noindent With this 'chirally improved' model we have repeated the
selfconsistent in-medium calculations described in the
previous section. The results are displayed in Fig.~3. We again
observe a reduction of the peak values in both $\sigma$- and
$\rho$- channel of the $\pi\pi$ scattering amplitude.
The effects are smaller than in the original J\"ulich model.
There remains some
enhancement over the vacuum curve in the low-energy region
and there is very little strength below threshold.
This can be attributed
to the subthreshold repulsion induced by the contact interactions.

\section{Summary}
Within the finite temperature Green's function method
{}~\cite{KaBa,SRS} we have derived
expressions for the pion selfenergy and
the uncorrelated 2$\pi$ propagator in a hot pion gas in thermal
as well as chemical equilibrium ($\mu_\pi =0$).
Starting from the vacuum $\pi\pi$ model
of Lohse et al.~\cite{LDHS} these have been used to calculate
medium modifications at the one- and two-body level
{\it selfconsistently}, {\it i.e.} the
in-medium scattering amplitude was evaluated in terms of the
pion selfenergy and vice versa.
It has been shown that the full off-shell calculations differ
from  the QPA in the following respects:
in the intermediate-energy range (E$\approx$300-800~MeV)
the imaginary part
of the $\pi\pi$ amplitude shows less suppression compared
to the vacuum curve; thermal excitation contributions (not
present in the QPA) lead to some  subthreshold
strength in the $\sigma$ channel at temperatures around
200~MeV; in particular, the QPA calculations overestimate
the thermal broadening and the upward peakshift of
the $\rho$ resonance.

\noindent Motivated by a recent analysis
of $\pi\pi$ correlations in cold nuclear matter~\cite{ARCSW},
we have also studied a 'chirally improved' version of the J\"ulich
$\pi\pi$ model. Applying this model to the hot pion gas
it is found that the subthreshold strength almost entirely
disappears due to the additional repulsion induced by
the chiral constraints. Also the enhancement just above the
two-pion threshold is much less pronounced.
The in-medium features of the $\pi\pi$ amplitude for
higher energies (reduction of the peak values
in both $\sigma$ and $\rho$ channel, essentially no
peakshift of the $\rho$ resonance) are in good agreement
for both models. After all, the deviations from the
$\pi\pi$ interaction in free space are rather small.

\noindent In conclusion, we wish to stress that proper off-shell
calculations beyond the QPA and the inclusion of constraints
from chiral symmetry are necessary  to obtain reliable
results for the pion selfenergy and the  in-medium $\pi\pi$
amplitudes of a hot pion gas. The latter could
serve as an input for transport models simulating the time
evolution of ultrarelativistic heavy-ion collisons~\cite{WeBe}
near freeze-out.

\vspace{2cm}
\bce
{\bf Acknowledgement}
\ece
\vspace{0.5cm}

\noindent
We thank  G.G. Bunatian, G. Chanfray, J. Durso, B. Friman and
J. Speth for useful discussions.
One of us (R.R.) acknowledges financial support by Deutscher
Akademischer Austauschdienst (DAAD).
This work was supported in part by a grant from the National Science
Foundation, NSF-PHY-89-21025.
\vfill\eject

\vfill\eject
\pagebreak
%
%
\bce
{\bf FIGURE CAPTIONS}
\ece
\vskip 2.0cm
\begin{itemize}
\item[{\bf Fig.~1}:]
The imaginary part of the $\pi\pi$ M-amplitude in $\sigma$ ($JI$=00)
and $\rho$ ($JI$=11) channel in a hot pion gas at $\mu_\pi$=0.
The long-dashed, short-dashed and dotted lines are the
selfconsistent results of the full off-shell calculations at
temperatures T=150,175 and 200~MeV, respectively. The dashed-dotted
line represents the selfconsistent QPA calculation at T=200~MeV.
The full line corresponds to the vacuum amplitude within the
non-chirally symmetric J\"ulich model.
\vskip 0.5cm

\item[{\bf Fig.~2}:]
A fit to the vacuum $\pi\pi$ scattering phase shift data obtained
within the chirally improved J\"ulich model.

\vskip 0.5cm

\item[{\bf Fig.~3}:]
The imaginary part of the $\pi\pi$ M-amplitude in $\sigma$ and
$\rho$ channel in a hot pion gas at $\mu_\pi$=0. All curves
are calculated by means of the chirally improved J\"ulich
model (full line: free space; temperature identification
as in fig.~1). The in-medium results are from selfconsistent,
full off-shell calculations as described in sect.~2.

\vskip 0.5cm

\end{itemize}

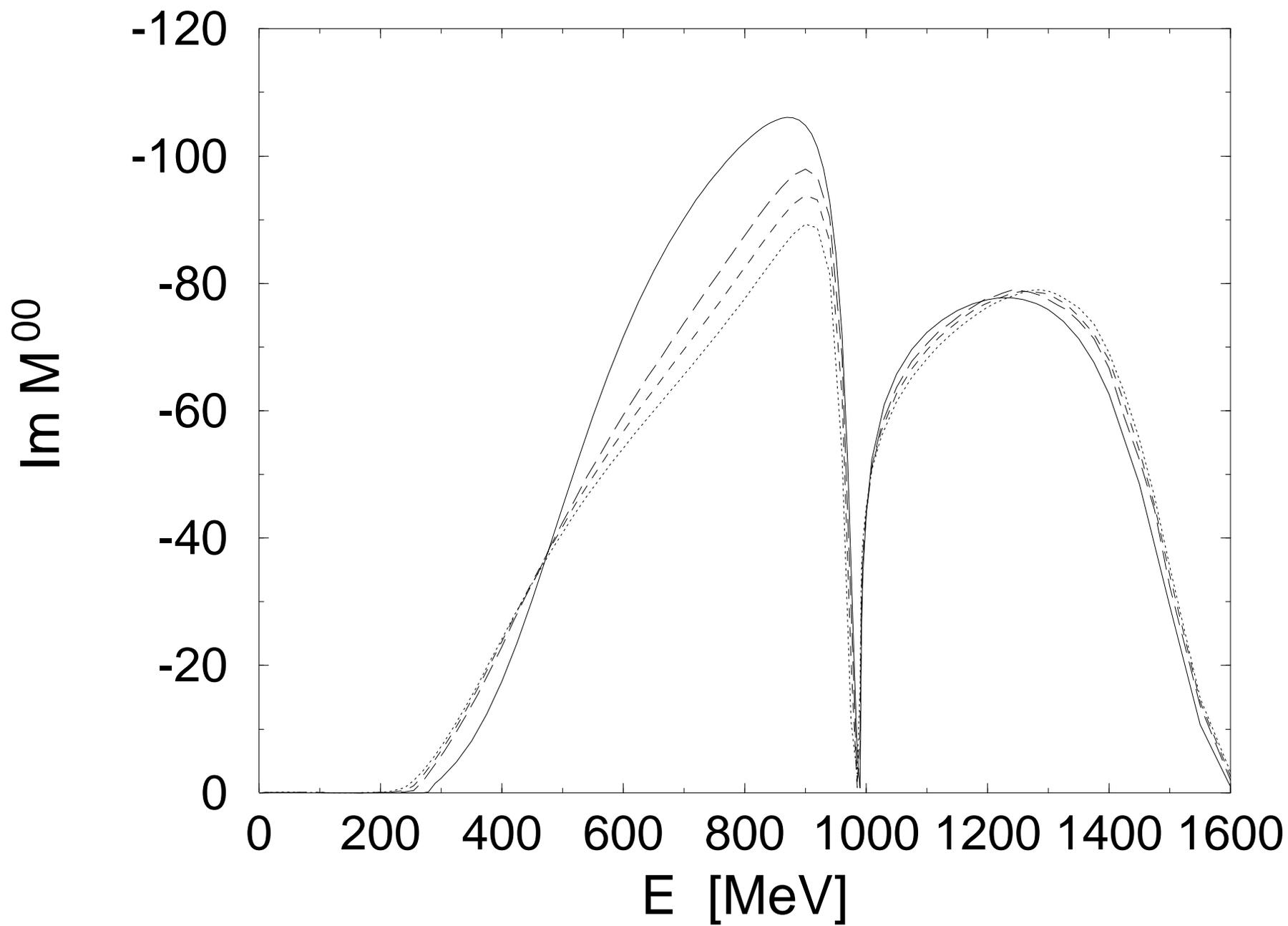
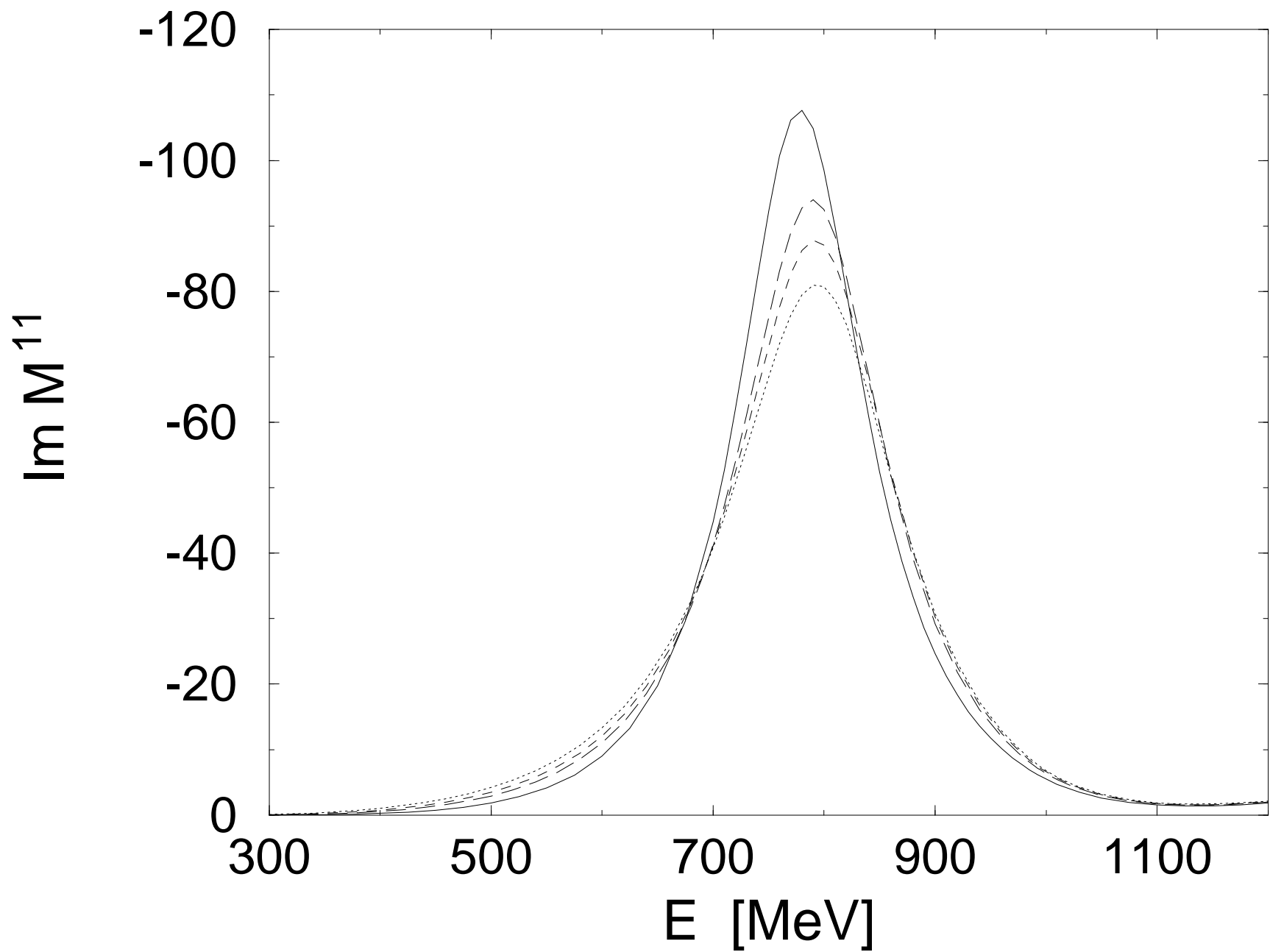
\end{document}